# Identification of a Marginal Fermi-Liquid in Itinerant Ferromagnet CoS$_2$

M. Otero-Leal,[1] F. Rivadulla,[2]*M. García-Hernández,[3] A. Piñeiro,[1,3] V. Pardo,[1,3] D. Baldomir,[1,3] J. Rivas[1]

*Departamentos de [1]Física Aplicada y [2]Química-Física, Universidad de Santiago de Compostela, 15782 Santiago de Compostela, Spain.*
[3]*Instituto de Investigaciones Tecnológicas, Universidad de Santiago de Compostela, 15782 Santiago de Compostela, Spain.*
[4]*Instituto de Ciencia de Materiales de Madrid (ICMM), CSIC, 28049-Cantoblanco, Madrid, Spain.*

**We report specific heat, resistivity and susceptibility measurements at different temperatures, magnetic fields, and pressures to provide solid evidence of CoS$_2$ being a marginal Fermi liquid. The presence of a tricritical point in the phase diagram of the system provides an opportunity to test the spin fluctuation theory with a high limit of accuracy. A magnetic field suppresses the amplitude of the spin fluctuations and recovers conventional Fermi liquid behavior, connecting both states continuously.**

The successful description of low temperature electronic properties of paramagnetic (PM) metals in terms of the Landau Fermi-Liquid (FL) model relies on the existence of fermionic quasiparticles, which stay long-lived provided their interactions remain short-ranged and repulsive. A strong exchange interaction among the quasiparticles can split the original PM Fermi surface into spin up/down bands, leading to a ferromagnetic (FM) instability in which the FL description is still valid. In the intermediate situation, *i.e.* the weak itinerant ferromagnet, the thermodynamic properties are governed by coupled long wavelength spin density fluctuations.[1] A quantitative treatment of the amplitude of the spin-density fluctuations and its temperature dependence is possible, and predicts key experimental observations like the Curie-Weiss susceptibility above T$_C$, or the enhanced paramagnetic moment, $\mu_{eff}$, with respect to the ordered moment, $\mu_s$ (the so called Rhodes-Wohlfarth ratio $\mu_{eff}/\mu_s$).[1,2] On the other hand, the long-wavelength character of the spin-fluctuations brings in long-range spin correlations that may well suppress the FL-like quasiparticle excitations in the itinerant weak ferromagnet. Actually, the low temperature properties of this new electronic state (referred to as the Marginal Fermi Liquid, MFL)[2] differ markedly from the FL state, as reflected for the 3D case in the behavior of the resistivity $\Delta\rho \sim T^{5/3}$, the magnetic susceptibility $\Delta\chi^{-1} \sim T^{4/3}$, or the specific heat $\Delta C/T \sim \ln T$. Some of these characteristics were observed in $Ni_xPd_{1-x}$ alloys in a narrow range of composition close to the FM limit,[3] and in weak ferromagnet $ZrZn_2$.[4] Other systems, like MnSi or $Sr_{1-x}Ca_xRuO_3$ show a more dramatic breakdown of the FL state, in which the magnetic phase is suppressed through a first-order phase transition and phase separation. [5,6,7] So, the MFL state is believed to provide the link between the conventional FL and other unconventional electronic states of matter, still to be properly defined.

Pyrite $CoS_2$ is a 3D FM metal whose $T_C$=122 K can be tuned to 0 K by application of pressure ($p_c \sim$ 60 kb).[8] We have measured a Rhodes- Wohlfarth ratio for this material $\mu_{eff}/\mu_s \sim 3.0$,[9] which places it in the weakly ferromagnetic limit in a situation close to that of MnSi.[1] On the other hand, different authors[8,10] suggested the proximity of this material to a tricritical point. From the thermal evolution of the magnetization under pressure, Goto *et al*.[11] placed this tricritical point at ~4 kb. Therefore, $CoS_2$ is a promising candidate to look for the realization of the MFL state.

Here we present high-field resistivity, specific heat and high-pressure magnetization results that demonstrate the MFL nature of the low temperature itinerant ferromagnet $CoS_2$. Beyond the tricritical point ($p_{tc}$~4.5(5) kb), a feature in the susceptibility points to a short-range magnetic order in a narrow temperature range $T_C < T < T^*$.

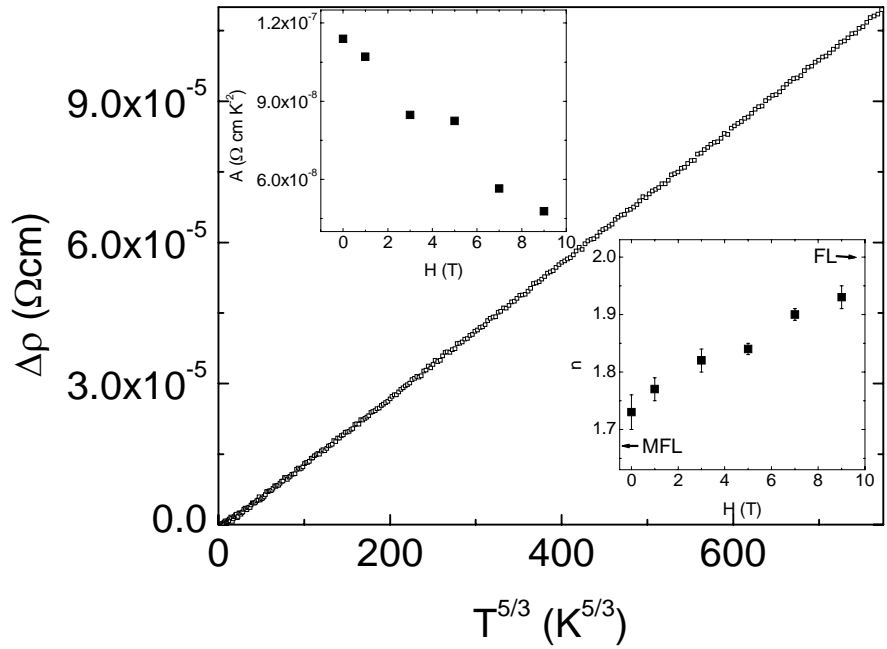

Figure 1: Resistivity vs. temperature according to the prediction of the spin fluctuation theory for a weak itinerant ferromagnet. The magnetic field dependence of the temperature exponent of the resistivity (lower inset) and the coefficient of the $T^n$ term (upper inset) in the resistivity ($\Delta\rho \sim AT^n$) are also show.

Polycrystalline $CoS_2$ was synthesized by conventional solid state reaction in evacuated silica tubes. Sulfur stoichiometry was ensured by thermo-gravimetrical analysis. In some cases the as-synthesized sample is non-stoichiometric and an additional thermal process is needed to reach the desired S/Co=2.00 ratio. Magnetization up to 10 kbar was measured in a commercial Be-Cu Cell from Easylab, using Sn as an internal manometer.

The residual resistivity of our samples was always between 40 to 50 μΩcm, about ten times higher than typical single crystals[12] but comparable to the best results in polycrystalline samples.[10] The results discussed below are independent of the residual resistivity and checked to be reproducible and sample independent.

In Fig. 1 we show the fitting of the resistivity to $\Delta\rho \sim T^{5/3}$. Careful log-log fitting to the equation $\Delta\rho \sim AT^n$ reveals that the exponent is actually slightly larger (~4%) than predicted by the spin fluctuation theory. On the other hand, a careful inspection of the log-log plot shows that below ~6 K, the exponent starts increasing somewhat pointing to the recovery of the FL phase at low temperatures, but measurements below 1.8 K should corroborate this point. In any case, the fitting to the $T^{5/3}$ law is satisfactory from 6 K up to ~40 K, demonstrating the persistence of the contribution of spin fluctuations to the resistivity. Above this temperature the resistivity increases faster than $T^{5/3}$ due to the influence of phonon scattering. Application of a magnetic field increases the resistivity exponent continuously towards the FL limit (lower inset to Fig.1) at the time that reduces the value of the coefficient of the $T^n$ term in the resistivity (upper inset to Fig.1).

Two groups previously reported a low temperature resistivity $\rho \sim AT^2$, consistent with a FL picture in this material (see references [8] and [10]). However, a careful inspection of the data in reference [10] show that the evolution of the resistivity is actually slightly slower than $T^2$ (note that we are proposing here a $\rho \sim AT^{1.7}$, difficult to distinguish from $T^2$ unless an accurate analysis of the transport is performed). On the other hand the fitting of the data in reference [8] is limited to a very narrow temperature range between 2K and 6K.

So, the results presented in Fig. 1 are consistent with a low temperature state in $CoS_2$ in which the transport properties are governed by long-wavelength spin fluctuations. It is also important to realize that suppression of the amplitude of the spin fluctuations by a magnetic field recovers the FL state and connects the MFL and the FL continuously. The results of the specific heat and magnetization that will be discussed below agree with this interpretation and confirm the intrinsic character of the effect observed in the resistivity, not related to the grain boundary scattering or any other source of disorder.

Exchange-enhanced spin fluctuations should manifest in an important enhancement of the electronic specific heat, which again should be sensitive to the application of a magnetic field. The results are shown in Fig. 2.

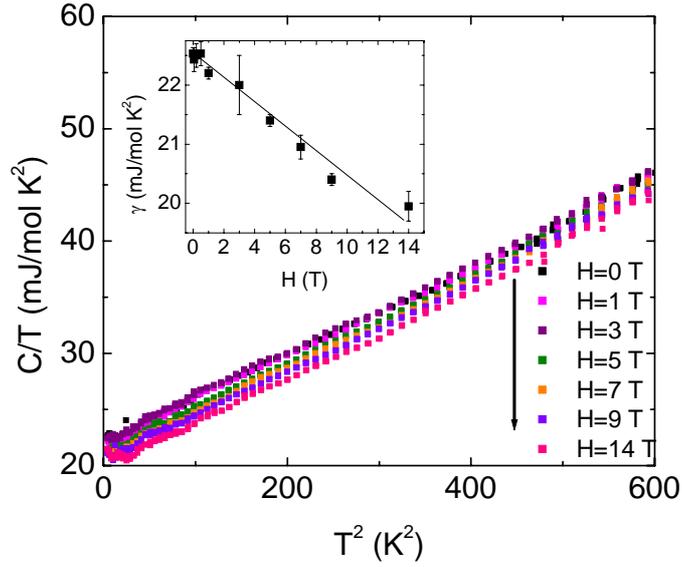

Figure 2: Specific heat at different magnetic fields. Inset: Evolution of γ with the magnetic field, as extracted from the fit of the data in the main plot. The reduction amounts to ~14% at 14 T.

The plot C/T vs $T^2$ at different fields gives a set of parallel straight lines between 10K and 30 K, resulting in a field independent Debye temperature $\theta_D$=352(2) K and a zero field value of the electronic specific heat coefficient γ = 22.5 (2) mJ/molK$^2$. We have performed ab initio calculations based on the density functional theory (DFT) using the WIEN2k software[13,14]. This uses a full-potential, all-electron scheme implemented via the APW+lo method[15]. The value obtained within the GGA approximation[16] for γ = 2.4 mJ/molK$^2$ is smaller than the experimental value by a factor 10. Our calculations do not include the spin fluctuation that could be key for such a large value of the parameter γ. Introducing strong correlation effects by using the LDA+U approach[17] does not improve the picture, reducing further the value obtained for γ (γ = 1.5 mJ/molK$^2$ for U~6 eV). Application of a magnetic field results in a considerable reduction of γ of about ~14% at 14 T (see the inset to Fig. 2).

On the other hand, although our data show a slight upturn at the lowest temperatures, the data are too noisy in this temperature region to discuss it on the basis of the ΔC/T~lnT behavior.

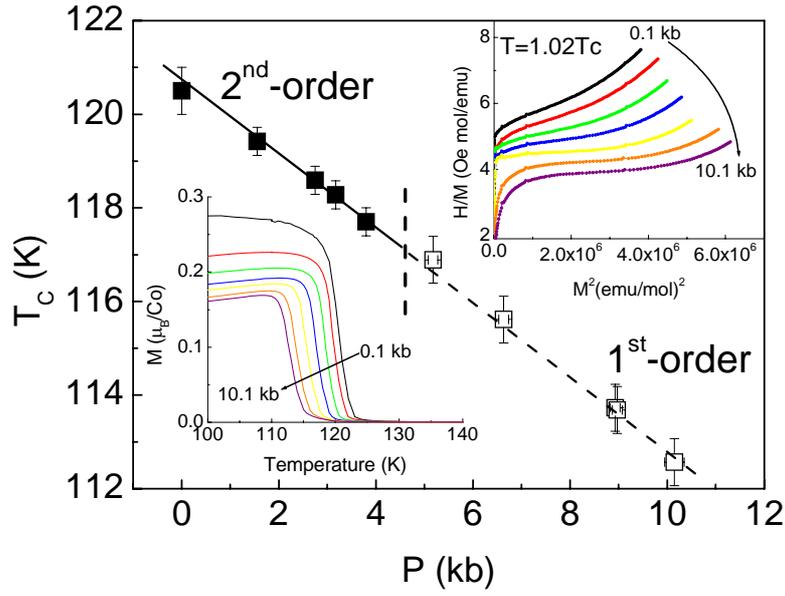

Figure 3: Pressure dependence of $T_C$. The vertical dashed line marks the transition from first-order to second-order magnetic phase transition, defining a tricritical point at $p_{tc} \sim 4.5\ (5)$ kb. Lower inset: Temperature dependence of the magnetization at different pressures. Upper inset: Pressure dependence of the H/M vs $M^2$ isotherm (T=1.02 $T_C$) for different pressures.

Now we will discuss the results of high-pressure magnetization at different fields and temperatures. From the determination of the pressure dependence of the magnetic susceptibility slightly above $T_C$ in this material we can verify the spin fluctuation theory to an unprecedented limit of accuracy. For an isotropic ferromagnet, $\chi$ can be derived from the expansion of the free energy in even terms of M:[18]

$$H = aM + bM^3 + cM^5 + \ldots \quad (1)$$

where the Pauli susceptibility, $\chi_0$, and the effective exchange field, $\lambda$, enter the equation in $a = (\chi_0 - \lambda)$. On the other hand, increasing temperature reduces the bulk FM moment of the system through the excitation of thermal fluctuations which determine the temperature dependence of the magnetization. This process can be treated by considering the effect of a random exchange field in equation (1), as shown by Lonzarich[2]

$$H = (a + 3b\, \bar{m}^2)M + bM^3 + \ldots \quad (2)$$

where $\bar{m}^2$ refers to the variance of the fluctuations. A derivation of the temperature dependence of $\bar{m}^2$ results in $\Delta\chi^{-1} \sim T^{4/3}$, for the case relevant here.[2] From equation (2) also follows that it is the existence of an anharmonic term (b≠0) that introduces the temperature dependence. If b→0, then higher order terms must be considered in the expansion (…+cM$^5$), and

$$A = a + 3b\,\bar{m}^2 + 5c(\bar{m}^2)^2 \text{ and } \Delta b = b + 10c\,\bar{m}^2 \qquad (3)$$

leading to a more rapidly varying susceptibility, $\Delta\chi^{-1} \sim (T^{4/3})^2$. Then, if the presence of a high-pressure first-order transition is confirmed in CoS$_2$, the system will go through a tricritical point characterized by b=0 in eq. (1), and hence the predictions of spin fluctuation theory could be tested with exceptional precision through the change in the rate of variation of $\Delta\chi^{-1}(T)$.

The pressure dependence of the magnetization is shown in Fig. 3 up to ~10 kb. $T_C$ decreases at an approximately linear rate of dln$T_C$/dP= 6.1(4) 10$^{-3}$ kb$^{-1}$. The sign of b in equation (1) and hence the nature of the magnetic phase transition can be obtained from the representation of the H/M vs M$^2$ isotherms, slightly above $T_C$.

The isotherms at different pressures at the same equivalent temperature (1.02 $T_C$) are shown in the inset of Fig. 3. The slope of the curves decreases with pressure, and from a fitting to equation (1), the transition becomes first-order (b<0) at ~4.5(5) kb. So, our measurements provide a direct demonstration of the existence of a tricritical point in the (P,T) phase diagram of CoS$_2$.

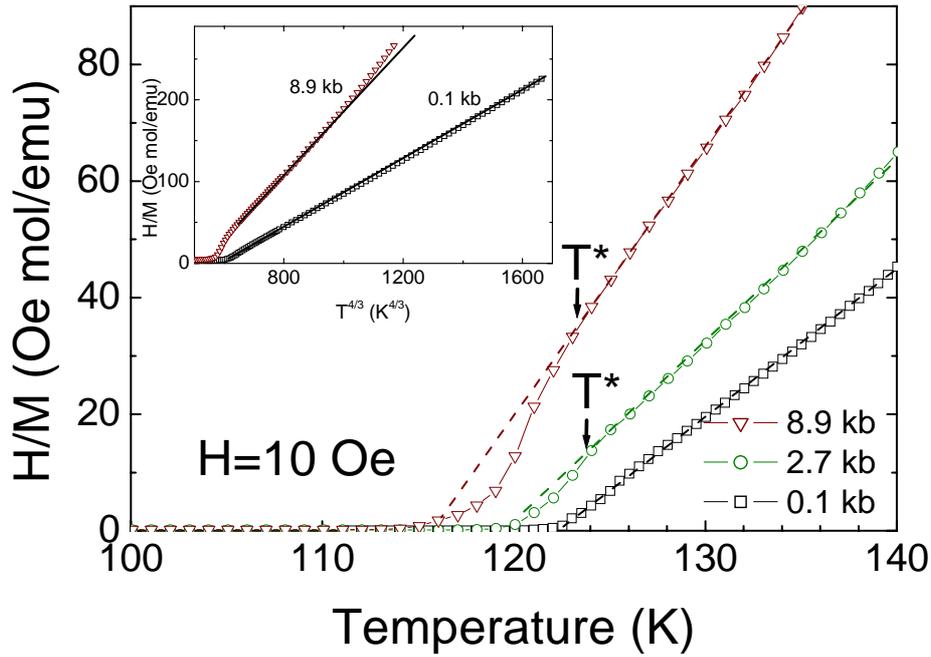

Figure 4: Low field (H=10 Oe) inverse paramagnetic susceptibility at different pressures, in the vicinity of $T_C$. The arrows mark the rapid downturn of the susceptibility below T* ($T_C$=120.5, 118.6 and 113.7 for 0.1, 2.7 and 8.9 kb, respectively). The apparent linear dependence of the inverse susceptibility above T* is due to the small temperature interval plotted. Inset: Inverse paramagnetic susceptibility (H=1T) at two different pressures, according to the expected $\Delta\chi^{-1}\sim T^{4/3}$ from the spin fluctuation theory. Increasing pressure results in a faster variation of the susceptibility, up to $\Delta\chi^{-1}\sim T^{2.13(1)}$. The straight lines are guides to the eye.

Following the previous discussion and equations (2) and (3), the inverse paramagnetic susceptibility should vary as $\sim T^{4/3}$ at low pressure, and increase the rate of variation up to $\sim (T^{4/3})^2$ at higher pressures. The results are shown in the inset to Fig. 4, fitted to the prediction of the theory. At ambient pressure $\Delta\chi^{-1}\sim T^{4/3}$, over 100 K above $T_C$. Actually, a closer fitting (log-log) of the ambient pressure susceptibility shows a $T^{1.48(1)}$, slightly larger (~10%) than predicted, probably from the contribution of higher order terms in the expansion. Increasing pressure increases the rate of variation, reaching $\Delta\chi^{-1}\sim T^{2.13(1)}$, providing a very accurate confirmation of spin fluctuation theory, and completes a consistent picture with the transport and specific heat measurements.

On the other hand, the low field susceptibility in the vicinity of $T_C$ is shown in Fig. 4. A sudden drop in the inverse susceptibility occurs in the interval $T_C < T < T^*$, becoming more evident at $p \geq p_{tc}$. The effect is completely erased by a relatively small magnetic field (H >1 kOe), and this could be the reason why it was not previously reported. On the other hand, the cooling and heating metastability limits derived from the fittings of the H/M vs $M^2$ curves around $T_C$ to equation (1), results in a predicted maximum hysteresis of ~0.1 K.[18] Consistent with this result, we have not observed thermal hysteresis in the M(T) curves.

A situation like that points to a rather weak first-order transition in $CoS_2$, in which the energy barrier separating the ordered and disordered states is strongly suppressed. In this scenario a thermodynamically favorable coexistence of the ordered and disordered phases becomes more plausible around $T_C$, like in the case of a spinodal phase segregation,[18] and the drop in the inverse susceptibility could be reflecting the existence of local magnetic order in the temperature interval $T_C < T < T^*$. This is further supported by a rapidly increasing $T_C$ with field, at a rate $d\ln T_C/dH = 3.3(2) \times 10^{-3}$ kOe$^{-1}$. Also consistent with our hypothesis is the report by Goto et al.[11] of a metamagnetic transition in a narrow range of temperature above the first-order magnetic phase transition.

On the basis of our results, the features of phase diagram of $CoS_2$ should be reconsidered at the light of the similarities shown with compounds like MnSi, and $ZrZn_2$, dominated by long-wavelength spin fluctuations. The existence of a tricritical point in these materials as $T_C$ is dropped to zero by pressure must not be accidental, but rooted in a common effect. Long-wavelength correlations are known to introduce a negative $M^4$ term in the expansion of the free energy (the equivalent to b in equation (1)), and hence the magnetic first-order phase transition should be generic to weak itinerant ferromagnets.[19] In the same line, spontaneous magnetic phase separation has been found to be common to many of these systems, as shown by Uemura et al.[7] The previous report of $\Delta\rho \sim T^{3/2}$ right beyond the quantum phase transition in $CoS_2$,[8] along with our report of phase separation, could point to a common origin of both effects.

In summary, the resistivity, specific heat and magnetic susceptibility results discussed above, demonstrate that the thermodynamic properties of $CoS_2$ are dominated by exchange-enhanced spin density fluctuations and provide a solid evidence for the occurrence of a MFL at low temperature in this material. Further experiments should confirm or exclude the $T^{3/2}$ phase beyond the pressure-induced quantum phase transition. If this is indeed confirmed, we will be in front of a (T, H, P) phase diagram in which the FL, MFL and $T^{3/2}$ phases will be accessible, and all that in a very simple system from the chemical, structural and preparative point of view.

**Acknowledgements**

Dr. M. Abd-Elmeguid and Dr. D. Khomskii are acknowledged by discussion and critical reading of the paper. We also thank Xunta de Galicia (Project No. PXIB20919PR) and the Ministry of Science of Spain (MAT2005-06024 and MAT2006-10027) for financial support. F.R. is also supported by the Ministry of Science of Spain under program Ramón y Cajal. We thank the CESGA (Centro de Supercomputación de Galicia) for the computing facilities.

[1] T. Moriya, in "Spin Fluctuations in Itinerant Electron Magnetism", Springer-Verlag, Berlin, 1985.

[2] G. G. Lonzarich, in "Electron", Ed. by M. Springford, Cambridge Univ. Press, Cambridge, 1997.

[3] M. Nicklas, M. Brando, G. Knebel, F. Mayr, W. Trinkl, A. Loidl, Phys. Rev. Lett. **82**, 4268 (1999).

[4] E. A. Yelland, S. J. Yates, O. Taylor, A. Griffiths, S. M. Hayden, A. Carrington, Phys. Rev. B **55**, 8330 (1997).

[5] C. Pfleiderer, G. J. McMullan, S. R. Julian, G. G. Lonzarich, Phys. Rev. B **55**, 8330 (1997).

[6] C. Pfleiderer, D. Reznik, L. Pintschovius, H. v. Löhneysen, M. Garst, A. Rosch, Nature **427**, 227 (2004).

[7] Y. J. Uemura et al. Nat. Phys. **3**, 29 (2007).

[8] S. Barakat et al. Phys. B **359-361**, 1216 (2005).

[9] We have measured a saturation moment $\mu_s \sim 0.9\mu_B/Co$. The effective moment $\mu_{eff} \sim 2.7$ $\mu_B/Co$ is only an approximation due to the non-linearity of the inverse susceptibility, as discussed in the text.

[10] L. Wang, T. Y. Chen, C. Leighton, Phys. Rev. B **69**, 094412 (2004).

[11] T. Goto, Y. Shindo, H. Takahashi, S. Ogawa, Phys. Rev. B **56**, 14019 (1997).

[12] K. Adachi, K. Ohkohchi, J. Phys. Soc. Japan **49**, 154, (1980).

[13] K. Schwarz and P. Blaha, Comput. Mater. Sci. **28**, 259 (2003).

[14] P. Blaha, K. Schwarz, G. K. H. Madsen, D. Kvasnicka, and J.Luitz, *WIEN2K, An Augmented Plane Wave Plus Local Orbitals Program for Calculating Crystal Properties*, ISBN 3-9501031-1-2 (Vienna University of Technology, Austria, 2001).

[15] E. Sjöstedt, L. Nördstrom, and D. J. Singh, Solid State Commun. **114**, 15 (2000).

[16] J. P. Perdew, K. Burke, and M. Ernzerhof, Phys. Rev. Lett. **77**, 3865 (1996).

[17] A. I. Liechtenstein, V. I. Anisimov, and J. Zaanen, Phys. Rev. B 52, R5467 (1995).

[18] P. M. Chaikin, T. C. Lubensky, in "Principles of Condensed Matter Physics" Cambridge University Press, Cambridge (1995).

[19] D. Belitz, T. R. Kirpatrick, T. Votja, Phys. Rev. Lett. **82**, 4707 (1999).